\documentclass{ws-ijmpa}

\begin{document}

\markboth{Probir Roy}
{The magic of four zero neutrino Yukawa textures}

%
\catchline{}{}{}{}{}
%

\title{The magic of four zero neutrino Yukawa textures
\footnote{Talk given at the ``CTP International Conference on Neutrino Physics in the LHC Era''(Luxor,Egypt, 15-19
Nov, 2009) and at ``Beyond the Standard Model, 2010''(Cape Town, South Africa, Feb 1-7, 2010)}}
\author{PROBIR ROY}

\address{DAE Raja Ramanna Fellow,\\
  Saha Institute of Nuclear Physics, 1/AF Bidhan
        Nagar, Kolkata 700064, India \\
probir.roy@saha.ac.in}

%

\maketitle


\begin{abstract}
Four is the maximum number of texture zeros allowed in the Yukawa coupling matrix of three massive neutrinos. These completely fix the
 high scale CP violation needed for leptogenesis in terms of  that accessible at laboratory energies. $\mu\tau$ symmetry drastically 
reduces such allowed textures. Only one form of the light neutrinos mass matrix survives comfortably while another is marginally allowed.

\end{abstract}

\ccode{PACS number(s): 14.60.Pq, 11.30.Hv, 98.80.Cq}

\section{Introduction}	

There is something magical about four zero Yukawa textures. After earlier success in the quark sector\cite{rev},
 it is proving useful with leptons, as explained in the abstract. Within the type-I seesaw framework and in the 
weak basis of mass diagonal charged leptons and heavy right chiral neutrinos, the results stated above were 
derived\cite{Branco:2007nb} leading to 
a highly constrained and predictive scheme\cite{Branco:2007nb,ctb}. We shall discuss the effect of the imposition of $\mu\tau$ symmetry\cite{Har} 
on this scheme.

The light neutrino mass matrix in the usual notation is 
\begin{eqnarray}
M_\nu\simeq-m_DM_R^{-1}m_D^T,
\end{eqnarray}
with $O(M_R)\gg O(m_D)$. $M_\nu$ diagonalizes as under
\begin{eqnarray}
U^\dagger M_\nu U^*=M^\nu_d={\rm
    diag} (m_1,~m_2,~m_3),
\end{eqnarray}
$m_{1,2,3}$ being real and positive. Our PMNS parametrization is 
\begin{eqnarray}
U=\left(\begin{array}{ccc}                     
1 & 0 & 0 \\
0 & c_{23} & s_{23} \\
0 & -s_{23} & c_{23} \end{array}\right)\left(\begin{array}{ccc}                     
c_{13} & 0 & -s_{13}e^{-i\delta_D} \\
0 & 1 & 0 \\
s_{13}e^{i\delta_D} & 0 & c_{13}\end{array}\right)\left(\begin{array}{ccc}                     
c_{12} & s_{12} & 0 \\
-s_{12} 
& c_{12}& 0 \\
0
&0 &1\end{array}\right)\left(\begin{array}{ccc} e^{i\alpha_M} & 0 & 0 \\ 0
& e^{i\beta_M} & 0 \\ 0 & 0 & 1\end{array}\right)
\end{eqnarray}
where $c_{ij} \equiv \cos\theta_{ij},~s_{ij}
\equiv \sin\theta_{ij}$ and $\delta_D$,
$\alpha_M,~\beta_M$  are the one Dirac phase and two Majorana phases respectively.
 In our basis, $M_{\ell}={\rm diag} (m_e,~ m_\mu,~m_\tau)$ and $M_R={\rm diag} (M_1,~M_2,~M_3)$, 
all mass eigenvalues being real and positive. The Casas-Ibarra \cite{Cas} form 
for $m_D$ which equals the neutrino Yukawa coupling matrix times the relevant Higgs VEV, is
\begin{eqnarray}
\label{iba}
m_D=iU\sqrt{M_\nu^{d}}R\sqrt{M_R},
\end{eqnarray}
where $R$ in general is an unknown complex orthogonal matrix: $R^TR=RR^T=I$. The best fit experimental numbers, needed by us, appear in Table 1.
Loosely, $R=\frac{\Delta m_{21}^2}{\vert\Delta m_{32}^2\vert}\simeq 3.2 \times 10^{-2}$, $\theta_{23}\simeq\frac{\pi}{4},~\theta_{12}\simeq\sin^{-1}\frac{1}{\sqrt{3}}$ and $\theta_{13}$ is small. 
We assume no massless neutrino, i.e. ${\rm det}M_\nu \ne 0$.

\begin{table}[ph]
\tbl{Best-fit experimental numbers from [6]}
{\begin{tabular}{@{}cc@{}} \toprule
Quantity & Experimental values \\\colrule
$\Delta m^2_{21}=m_2^2-m_1^2$& $7.59\pm 0.20\left(^{+0.61}_{-0.69}\right)\times 10^{-5}$ $eV^2$ \\ 
&\\
 $\Delta m^2_{32}=m_3^2-m_2^2$& $-2.40\pm 0.11\left(^{+0.37}_{-0.36}\right)\times 10^{-3}$ $eV^2$(inverted) \\
&\\

& $2.51\pm 0.12\left(^{+0.39}_{-0.36}\right)\times 10^{-3}$ $eV^2$(normal)\\
&\\
$\theta_{12}$ & ${34.4\pm 1.0\left(^{+3.2}_{-2.9}\right)}^\circ$ \\
&\\
$\theta_{23}$ & ${42.3^{+5.3}_{-2.8}\left(^{11.4}_{-7.1}\right)}^\circ$ \\
&\\
$\theta_{13}$ & $<13.2^\circ$ \\ \botrule
\end{tabular} }
\end{table}
\section{Four zero Yukawa textures and $\mu\tau$ symmetry}
It is more natural to attribute ab initio textures to $m_D$, which appears in the Lagrangian rather than to the derived $m_\nu$. $72$ allowed four zero textures in $m_D$ 
have been classified into\cite{Branco:2007nb} two categories: (A) $54$ with one pair of vanishing conjugate off diagonal elements in $M_\nu$ and 
(B) $18$ with two zeros in one row and one each in the other two ($k,l$ say) obeying 
${\rm det}({\rm cofactor} (M_\nu)_{kl})=0$. For all these, the $R$ matrix of Eq. (\ref{iba}) has been reconstructed in terms of the 
element of $U$, $M_\nu^{\rm d}$ and $M_R$. Consequently, all phases of $R$ are given in terms
of $\delta_D$,
$\alpha_M$ and $\beta_M$ which completely determine all phases in $m_D$ including those responsible
for leptogenesis.

Elements of $m_D$ and $M_R$ are required by $\mu\tau$ symmetry to
remain invariant under the interchange $\nu_\mu\leftrightarrow$
$\nu_\tau$, $N_\mu\leftrightarrow N_\tau$.  The seesaw formula
immediately implies a custodial $\mu\tau$ symmetry in $M_\nu$ itself,
leading to $\theta_{23}=\frac{\pi}{4}$, $\theta_{13}=0$. Further, the
number of four zero textures allowed in $m_D$ is drastically
reduced\cite{Adh}. Only two each are allowed in categories (A) and (B) both
leading to the same $M_\nu^{(A)}$ or $M_\nu^{(B)}$:
\begin{eqnarray}
\label{numatab}
M_\nu^{(A)} = m \left(\begin{array}{ccc} k_1^2e^{2i\alpha}+2k_2^2&k_2&k_2\\
                                    k_2&1&0\\
                                    k_2&0&1\end{array}\right) \qquad M_\nu^{(B)} = m'\left(\begin{array}{ccc} l_1^2&l_1l_2e^{i\beta}&l_1l_2e^{i\beta}\\
                                    l_1l_2e^{i\beta}&l_2^2e^{2i\beta}+1&l_2^2e^{2i\beta}\\
                                    l_1l_2e^{i\beta}&l_2^2e^{2i\beta}&l_2^2e^{2i\beta}+1\end{array}\right).
\end{eqnarray}
Here $m$ and $m'$ are overall mass scales, $k_1,~k_2,~l_1,~l_2$ are real
parameters and $\alpha$, $\beta$ are phases.

Turning to $\theta_{12}$ and $R$, one can derive that
\begin{eqnarray}
\label{rt12}
&&R = 2{(X_1^2+X_2^2)}^{1/2}{[X_3-{(X_1^2+X_2^2)}^{1/2}]}^{-1},\\\nonumber
&&\tan2\theta_{12} = \frac{X_1}{X_2}.
\end{eqnarray}
The $X$'s of Eq. (\ref{rt12}) are given for $M_\nu^{(A)}$ as
\begin{eqnarray}
\label{xa}
X_1^{(A)} &=& 2\sqrt{2}k_2{[{(1+2k_2^2)}^2 + k_1^4 + 2k_1^2(1+2k_2^2)\cos2\alpha]}^{1/2},\\\nonumber
X_2^{(B)} &=& 1-k_1^4-4k_2^4-4k_1^2k_2^2\cos2\alpha,\\\nonumber
X_3^{(A)} &=& 1-4k_2^4-k_1^4-4k_1^2k_2^2\cos2\alpha - 4k_2^2. 
\end{eqnarray}
For $M_\nu^{(B)}$, they are given by 
\begin{eqnarray}
\label{xb}  
X_1^{(B)} &=& 2\sqrt{2}l_1l_2{[{(l_1^2+2l_2^2)}^2 + 1+ 2(l_1^2+2l_2^2)\cos2\beta]}^{1/2},\\\nonumber
X_2^{(B)} &=& 1+4l_2^2\cos2\beta+4l_2^4-l_1^4,\\\nonumber
X_3^{(B)} &=& 1-{(l_1^2+2l_2^2)}^2 - 4l_2^2\cos2\beta.
\end{eqnarray}

One can further impose the requirement of tribimaximal mixing,
i.e. $\theta_{12}=\sin^{-1}\frac{1}{\sqrt{3}}\simeq 35.2^\circ$, which
needs $(M_\nu)_{11}+(M_\nu)_{12}=(M_\nu)_{22}+(M_\nu)_{23}$. For
$M_\nu^{(A)}$, $\alpha$ is then immediately fixed at $\pi/2$ and
$k_1=(2k_2^2+k_2-1)^{1/2}$. For $M_\nu^{(B)}$, $\beta$ is then
immediately fixed at $\cos^{-1}(l_1/4l_2)$ and
$l_2=(1-l_1^2)^{1/2}/2$. For category (A) $R$ equals
$3(k_2-2)/(k_2+1)$, while for category (B) it becomes
$3l_1^2/2(1-2l_1^2)$.
\begin{figure}[pt]
\centerline{\psfig{file=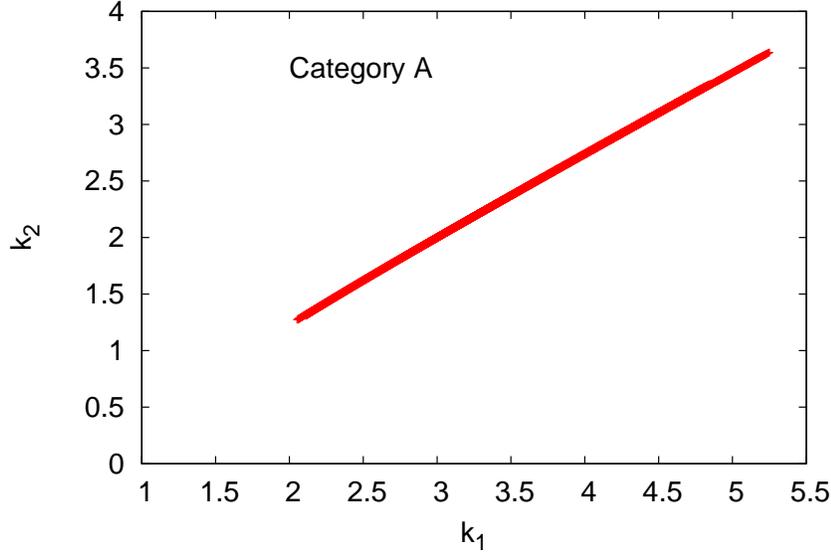,width=8cm,angle=270}}
\vspace*{8pt}
\caption{Variation of $k_1$ and $k_2$ in category A with $\mu\tau$ symmetry over the $3\sigma$ allowed ranges of 
$R$ and $\theta_{12}$.  
 \label{f1}}
\end{figure}
\section{Phenomenology}   
Experimentally fitted values of $R$ and $\tan 2\theta_{12}$ can be
matched with the expression in Eq. (\ref{rt12}) - Eq. (\ref{xb}). For category (A), only the
spectrum with inverted ordering is found to be allowed. But no common
allowed parameter ranges are found for ($k_1,~k_2,~\alpha$) in the $1\sigma$ intervals of $R=-(2.88-3.37)\times 10^{-2}$ and $\theta_{12}=33.15^\circ-35.91^\circ$. The $3\sigma$ intervals $R=-(2.46-3.99)\times 10^{-2}$ and $\theta_{12}=30.66^\circ-39.23^\circ$ allow a thin strip (Fig. 1) in the $k_1-k_2$ plane with $89^\circ\le \alpha \le 90^\circ$ and $2.0<k_1<5.3$, $1.2<k_2<3.7$. The additional constraint of tribimaximal mixing, when $\alpha$ must be exactly $\pi/2$, confines $k_2$ to the range $1.95\le k_2\le 1.97$. Improved experimental errors may thus rule out  this category completely.


Only the normally ordered spectrum is found to be allowed for category (B).
But now there are significant allowed regions in the $l_1-l_2$ plane both for $1\sigma$ and $3\sigma$ intervals $R$ and $\theta_{12}$. 
Specifically for the  $3\sigma$ intervals $R=(2.52-4.07)\times 10^{-2}$ and $\theta_{12}=30.66^\circ-39.23^\circ$, two allowed branches 
appear (Fig.2) with $\beta$ in the ranges $87^\circ$ to $90^\circ$, 
$0.1<l_1<0.55$ and $0.6<l_2<0.76$. The further imposition of
tribimaximal mixing forces the only one free variable left, namely
$l_1$, to be within the range $0.11\le l_1\le 0.15$.

\begin{figure}[pb]
\centerline{\psfig{file=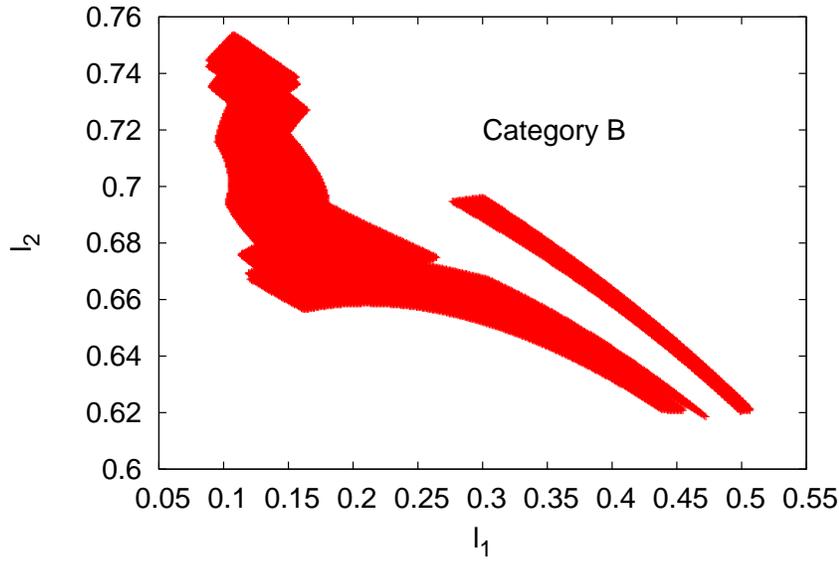,width=8cm,angle=270}}
\vspace*{8pt}
\caption{Variation of $l_1$ and $l_2$ in Category B for the $3\sigma$ allowed ranges of 
$R$ and $\theta_{12}$ \label{f2}}
\end{figure}
%
\section{Radiative lepton flavor violating decay and leptogenesis}
With $\alpha>\beta$ and $l_1 = l_e, l_2 = l_\mu, l_3 = l_\tau$, the branching ratio for the decay $l_\alpha\rightarrow l_\beta\gamma$ can be written in mSUGRA
scenarios (with universal scalar masses at a GUT scale $M_\chi\sim 2\times10^{16}$ GeV) as
\begin{eqnarray}
{\rm{BR}}(l_\alpha\rightarrow l_\beta\gamma)\propto{\rm BR}(l_\alpha\rightarrow l_\beta\nu{\bar\nu})
|(m_DLm_D^\dagger)_{\alpha\beta}| 
\end{eqnarray}
with 
\begin{eqnarray}
L_{kl} = \ln\frac{M_X}{M_k}\delta_{kl},
\end{eqnarray}
$M_k$ being the mass of the $k$th heavy right chiral neutrino $N_k$.  Now ${\rm{BR}}(\tau\rightarrow \mu\gamma)$ vanishes for category (A) since
$(M_\nu^{(A)})_{23}=0$. Otherwise, the allowed textures in both categories have $(M_\nu)_{13}=(M_\nu)_{12}\ne0$ and lead to nonzero
${\rm{BR}}(\tau\rightarrow e\gamma)$ and ${\rm{BR}}(\mu\rightarrow e\gamma)$ but with the relation
\begin{eqnarray}
\frac{{\rm BR}(\tau\rightarrow e \gamma)}{{\rm BR}(\mu\rightarrow e \gamma)}\simeq 
\frac{\rm{BR}(\tau\rightarrow e\nu_e\bar\nu_e)}{\rm{BR}(\mu\rightarrow e\nu_\mu\bar\nu_\mu)} \simeq 0.178.
\end{eqnarray}
For leptogenesis, the flavor dependent lepton asymmetry in the standard notation is given for the Minimal Supersymmetric Standard Model by
\begin{eqnarray}
\varepsilon_i^\alpha &=& \frac{\Gamma(N_i\rightarrow \phi\bar{l}_{\alpha})-\Gamma(N_i\rightarrow\phi^\dagger l_\alpha)}
{\sum_{\beta} [\Gamma(N_i\rightarrow\phi\bar{l_\beta}) + \Gamma(N_i\rightarrow\phi^\dagger l_\beta)]}
\nonumber\\
&\simeq& 
\frac{g^2}{16\pi M_W^2}\frac{1}{{(m_D^\dagger m_D)}_{ii}}\sum_{j\neq i}\left[{\cal I}^\alpha_{ij} 
f\left(\frac{M_j^2}{M_i^2}\right) + {\cal J}^\alpha_{ij}{\left(1- \frac{M_j^2}{M_i^2}
\right)}^{-1}
\right], 
\end{eqnarray}
\begin{eqnarray}
{\cal I}^\alpha_{ij} = {\rm{Im}}{(m_D^\dagger)}_{i\alpha}{(m_D)}_{\alpha j}{(m_D^\dagger m_D)}_{ij} 
= -{\cal I}^\alpha_{ji},\\
{\cal J} ^\alpha_{ij} = {\rm{Im}} 
{(m_D^\dagger)}_{i\alpha}{(m_D)}_{\alpha j}{(m_D^\dagger m_D)}_{ji} = 
-{\cal J}^\alpha_{ji},\\
f(x) = \sqrt{x}\left[\frac{2}{1-x} - \ln\frac{1+x}{x}\right].
\end{eqnarray}

The flavor independent lepton asymmetry is
\begin{eqnarray}
\varepsilon_i = \sum_\alpha \varepsilon_i^\alpha = \frac{g^2}{16\pi M_W^2}\frac{1}{{(m_D^\dagger m_D)}_{ii}} 
\sum_{j\neq i}{[{(m_D^\dagger m_D)}_{ij}]}^2 f\left(M_j^2/M_i^2\right).
\end{eqnarray}
For $M_1<< M_{2,3}$, $f(M_{2,3}^2/M_1^2)\simeq -3 M_1/M_{2,3}$. Table 2. summarizes our statement on the leptogenesis parameters including
the effective mass $\widetilde{m_1}^\alpha=|(m_D)_{\alpha 1}|^2/M_1$ for the washout of the $\alpha$-flavor asymmetry
\begin{table}[pt]
\tbl{Leptogenesis Table}
{\begin{tabular}{@{}cccccc@{}} \toprule
configuration & ${\cal I}_{ij}^\alpha$& ${\cal J}_{ij}^\alpha$&$\widetilde{m_1}^e$
&$\widetilde{m_1}^\mu$&
$\widetilde{m_1}^\tau$\\\colrule
$m_D^{(1)}$ & ${\cal I}^e_{12}={\cal I}^e_{13}\neq 0$, \rm{rest zero}&0&nonzero&0&0\\
&&&&&\\
$m_D^{(2)}$ & --do-- & 0&nonzero&0&0\\
&&&&&\\
$m_D^{(3)}$ & ${\cal I}^\mu_{12}={\cal I}^\tau_{13}\neq 0$, \rm{rest zero}&0&nonzero&nonzero&equals 
$\widetilde{m_1}^\mu$\\
&&&&&\\
$m_D^{(4)}$ & ${\cal I}^\mu_{13}={\cal I}^\tau_{12}\neq 0$, \rm{rest zero}& 0 &nonzero &nonzero &equals 
$\widetilde{m_1}^\mu$\\\\ \botrule
\end{tabular} \label{ta2}}
\end{table}
\section{Deviation due to RG running}
Suppose $\mu\tau$ symmetry in the neutrino sector is imposed at a high scale $\Lambda\sim 10^{12}$ GeV. The neutrino mass matrix elements can then be evolved down to a 
laboratory scale $\lambda\sim 10^3$ GeV by RG running. On account of the inequality $m_\tau\gg m_\mu\gg m_e$, $\mu\tau$ symmetry is badly broken in the charged lepton sector. Deviations from $\mu\tau$ symmetry creep into
$M_\nu$ from loop diagrams with charged lepton internal lines. We keep only $m_\tau$ induced terms via $\Delta_\tau \simeq \frac{m_\tau^2}{8\pi^2 v^2}{(\tan^2\beta +1)}\ln\left(\frac{\Lambda}{\lambda}
\right)$ with $v^2=v_u^2+v_d^2$ and $\tan\beta=v_u/v_d$, $v_{u,d}$ being up, down type Higgs VEVs in the MSSM. 
Then to $O(\Delta_\tau)$, $(M_\nu)_{11}$, $(M_\nu)_{12}$, $(M_\nu)_{21}$ and $(M_\nu)_{22}$ are unchanged but the remaining elements change to 
$(1-\Delta_\tau)(M_\nu)_{13}$, $(1-\Delta_\tau)(M_\nu)_{31}$, $(1-\Delta_\tau)(M_\nu)_{23}$, $(1-\Delta_\tau)(M_\nu)_{32}$ and 
$(1-2\Delta_\tau)(M_\nu)_{33}$. Consequently, $\theta_{13}$ can be nonzero and $\theta_{23}$ different from $\pi/4$.

One can redo the phenomenology with these changes. For category (A), the $3\sigma$ allowed strip in the $k_1-k_2$ plane gets marginally extended now with
$0\le\theta_{13}^\lambda\le 2.7^\circ$, $\theta_{23}\le 45^\circ$ while the inverted ordering is retained and the normal one excluded. For category (B),
the $3\sigma$ allowed branches in the $l_1-l_2$ plane are enhanced a bit more with $0\le\theta_{13}^\lambda\le 0.85^\circ$, $\theta_{23}\ge 45^\circ$
while retaining the normal ordering and excluding the inverted one.
\section{Conclusion}
\begin{enumerate}
\item Just four neutrino Yukawa textures with four zeros are compatible with $\mu\tau$ symmetry, leading to only two forms for the light neutrino mass matrix $M_\nu^{(A)}$ and $M_\nu^{(B)}$.
\item For $M_\nu^{(A,B)}$, $3\sigma$-allowed values of $\theta_{12}$
  and $R=\Delta m_{21}^2/\Delta m_{32}^2$ admit restricted regions in the parameter space with
  $M_\nu^{(A)}$ being in some tension with data
\item The tribimaximal mixing assumption further restricts the the parameters.
\item Radiative deviations from $\mu\tau$ symmetry yield small values
  of $\theta_{13}$ and can resolve the $\theta_{23}$ octant ambiguity
  ($<$ or $> 45^\circ $)

\end{enumerate}

\section*{Acknowledgments}

I thank K.S. Babu and my collaborators B. Adhikary and A. Ghosal for helpful discussions. This work has been supported by a DAE Raja Ramanna fellowship.

\end{document}